\documentclass{aa}
\usepackage{graphicx}
\usepackage{txfonts}
\usepackage{natbib}
\bibpunct{(}{)}{;}{a}{}{,}

\usepackage{lscape}
\usepackage{longtable}
\usepackage{amssymb}
\def\kms{km~s$^{-1}$}
\def\cep{Cepheus\,A\,HW2}

\usepackage{color}

\begin{document}

% Version 1
%   \title{The infalling and magnetized ring around Cepheus\,A\,HW2}
%   \subtitle{...}
%
% Version 2
   \title{Planar infall of CH$_3$OH gas around Cepheus\,A\,HW2}
%   \subtitle{...}
%
% Version 3
%   \title{}
%   \subtitle{...}

   \author{A. Sanna\inst{1} \and L. Moscadelli\inst{2} \and G. Surcis\inst{3,4} \and H.\,J. van Langevelde\inst{4}
               \and  K.\,J.\,E. Torstensson\inst{5}  \and A.\,M. Sobolev\inst{6}}
            
%\fnmsep\thanks{Just to show the usage of the elements in the author field}

   \institute{Max-Planck-Institut f\"{u}r Radioastronomie, Auf dem H\"{u}gel 69, 53121 Bonn, Germany\\
   \email{asanna@mpifr-bonn.mpg.de}
   \and INAF, Osservatorio Astrofisico di Arcetri, Largo E. Fermi 5, 50125 Firenze, Italy
   \and INAF, Osservatorio Astronomico di Cagliari, via della Scienza 5, 09047, Selargius, Italy
   \and JIVE, Joint Institute for VLBI in Europe, Postbus 2, 7990 AA Dwingeloo, The Netherlands
   \and European Southern Observatory, Alonso de Cordova 3107, Casilla 19001, Santiago 19, Chile
   \and Ural Federal University, 51 Lenin Str., Ekaterinburg 620000, Russia}

%  W.H.T. Vlemmings: \and Department of Earth and Space Sciences, Chalmers University of Technology, Onsala Space Observatory, SE-439 92 Onsala, Sweden

   \date{Received ...; accepted ...}

% \abstract{}{}{}{}{}
% 5 {} token are mandatory

  \abstract
   {}
   {In order to test the nature of an (accretion) disk in the vicinity of Cepheus\,A\,HW2, we measured the three-dimensional velocity 
   field of the CH$_3$OH maser spots, which are projected within 1000\,au of the HW2 object, with an accuracy of the 
   order of 0.1\,km\,s$^{-1}$.}
   {We made use of the European VLBI Network (EVN) to image the 6.7\,GHz CH$_3$OH maser emission towards Cepheus\,A\,HW2
   with 4.5\,milli-arcsecond resolution (3\,au). We observed at three epochs spaced by one year between 2013 and 2015. During the
   last epoch, on mid-march 2015, we benefited from the new deployed Sardinia Radio Telescope.}
   {We show that the CH$_3$OH velocity vectors lie on a preferential plane for the gas motion with only small deviations
   of $12\degr\pm9\degr$ away from the plane. This plane is oriented at a position angle of $134\degr$ east of north, and inclined
   by $26\degr$ with the line-of-sight, closely matching the orientation of the disk-like structure previously reported by \citet{Patel2005}.
   Knowing the orientation of the equatorial plane, we can reconstruct a face-on view of the CH$_3$OH gas kinematics
   onto the plane. CH$_3$OH maser emission is detected within a radius of 900\,au from HW2, and down to a radius of about 300\,au,
   the latter coincident with the extent of the dust emission at 0.9\,mm. The velocity field is dominated by an infall component of about
   2\,km\,s$^{-1}$  down to a radius of 300\,au, where a rotational component of 4\,km\,s$^{-1}$ becomes dominant. We discuss the
   nature of this velocity field and the implications for the enclosed mass.}
   {These findings bring direct support to the interpretation that the high-density gas and dust emission, surrounding
   Cepheus\,A\,HW2, trace an accretion disk.}

%  -- Techniques: high angular resolution

   \maketitle
%
%________________________________________________________________

\section{Introduction}

Cepheus\,A is the second nearest high-mass star-forming region (after Orion), located at a trigonometric distance of 700\,pc from
the Sun \citep{Moscadelli2009,Dzib2011} in the Cepheus\,OB3 complex (e.g., Fig.\,27 of \citealt{Kun2008}). The region has an IR
luminosity in the range 2--3\,$\rm\times10^4\,L_{\odot}$ (e.g., \citealt{Mueller2002,DeBuizer2016}) and hosts a tight
($<$\,7000\,au) cluster of young stellar objects (YSOs) associated with radio continuum emission. Half of the bolometric luminosity
is attributed to the brightest radio source in the field, named  ``HW2'' after \citet[see also \citealt{Garay1996}]{Hughes1984}, which
would correspond to a ZAMS star with an early-B  spectral type and a mass in excess of 10\,M$_{\odot}$. The HW2 object drives one
of the best examples of radio thermal jets in the literature (e.g., \citealt{Rodriguez1994,Curiel2006}), which excites several shocked layers 
of H$_2$O maser emission (e.g., \citealt{Torrelles2011}).
At the origin of the radio jet, \citet{Patel2005} resolved a dense core of dust and gas in the 345\,GHz band of the SMA (see
also \citealt{Torrelles2007}). The core emission is flattened in the direction perpendicular to the radio jet, with an aspect ratio near 2 between 
its major and minor axes. Assuming the core is tracing a circumstellar disk, \citeauthor{Patel2005} showed that the aspect ratio of the core 
emission implies a disk inclination of $28\degr$ with the line-of-sight. On the plane of the sky, the major axis of the disk, with an outer radius larger than
$0.8''$ (or 560\,au), is oriented at a position angle (P.A.) of $121\degr$ (east of north). Despite the fact that this simple picture would support
a scenario of disk-mediated accretion onto HW2, the environment around HW2 is significantly more complex than that of an isolated YSO, showing
the presence of perhaps more than 5 objects within a radius of 1000\,au (e.g., Fig.\,1 of \citealt{JimenezSerra2009}). Because of this
multiplicity, the disk scenario has been questioned as due to a chance superposition of different hot cores (e.g., \citealt{Brogan2007,Comito2007}).

%_____________________________________________________________
%                                                    TABLES  # 1
%-----------------------------------------------------------------------------------------------------------

%\addtocounter{table}{0}
%\begin{landscape}
\begin{table*}
\caption{\label{tab1} Summary of EVN observations (code ES071)}
\centering
\begin{tabular}{c c c c c c c c c}

\hline \hline
& & & & &  \multicolumn{2}{c}{Absolute Position  ($\pm1$\,mas)}  & &  \\
 Epochs\tablefootmark{a} & run & $\nu_{\rm rest}$   & $\rm \Delta v$  &  $\rm V_{\rm LSR}$ & R.A.\,(J2000) &     Dec.\,(J2000)      &  HPBW &           rms              \\
      (gg/mm/aa)               & (hr) &       MHz           & (km\,s$^{-1}$) &     (km\,s$^{-1}$)     &      (h m s)     &   ($\degr$ $'$ $''$)  &  (mas) & (mJy\,beam$^{-1}$)   \\

\hline
& & & & & & & &  \\
% 2013\,Mar.\,1\,/\,2014\,Feb.\,28\,/\,2015\,Mar.\,13   & 8  & C  &  0.04 & --3.0 & 22:56:17.9051 & 62:01:49.584  &  4.5 & 4--6  \\ tracking center used
01/03/13 -- 28/02/14 -- 13/03/15   & 8  & 6668.5  &  0.04 & --1.73 & 22:56:17.9089 & 62:01:49.527  &  4.5 & 4--6  \\
%2014\,Feb.\,28   & & & & & & & & \\
%2015\,Mar.\,13   & & & & & & & & \\
%& & & & & & & &  \\
\hline
%& & & & & & & &  \\
\end{tabular}

\tablefoot{Columns\,1 and 2: observing dates and duration of each run. Columns\,3 and 4: rest frequency and velocity resolution of the maser lines.
Columns\,5, 6, and 7: local-standard-of-rest velocity and measured absolute position of the reference maser spot used for calibration. Column\,8: restoring
beam size (round) used at each epoch. Column\,9: thermal noise achieved per line-free channel map at each epoch (Stokes\,\rm{I}). 
\tablefoottext{a}{The EVN operated with 8 antennas at the first two epochs (EF, WB, JB, ON, MC, NT, TR, YS), with the addition of SR during the last epoch.}}
\end{table*}
%\end{landscape}

%_____________________________________________________________
%-----------------------------------------------------------------------------------------------------------

By directly measuring the velocity field of gas belonging to the putative disk, we can provide a critical test to the disk-like scenario.  
Compact maser emission centers, of the order of a few au in size, are ideal test particles to probe the local, three-dimensional, gas
kinematics, by combining their proper motion vectors (i.e., their displacement on the plane of the sky) with the Doppler shift of the
maser lines (e.g., \citealt{Moscadelli2011,Torrelles2011}). Towards Cepheus\,A, bright CH$_3$OH maser emission at 6.7\,GHz was firstly reported by
\citet{Menten1991} and imaged at milli-arcsecond resolution by \citet{Sugiyama2008b,Sugiyama2008a} and \citet{Torstensson2008}.
Individual CH$_3$OH masers are projected within 1000\,au of the HW2 object, and outline a filamentary distribution with LSR velocities
redshifted, by less than 5\,km\,s$^{-1}$, with respect to the systemic velocity of the region (around  $-4.5$\,km\,s$^{-1}$). Following
this evidence, it has been proposed that the CH$_3$OH masers are tracing a contracting circular ring with radius of about 680\,au,
centered on HW2, and oriented nearly edge-on with respect to the observer \citep{Torstensson2008,Vlemmings2010,Torstensson2011,Sugiyama2014}.
Based on the detection of an ordered polarization field through the CH$_3$OH maser emission, \citet{Vlemmings2010} also suggested 
that the maser motions might be driven by magnetic forces.

In order to test the disk-like scenario, we exploited the high sensitivity of the European VLBI Network (EVN) to observe at different epochs
the 6.7\,GHz CH$_3$OH masers in HW2, with the aim of eventually tracing proper motions with an accuracy
of 0.1\,km\,s$^{-1}$. We present the details of the EVN observations in Sect.\,\ref{obs}. In Sect.\,\ref{results}, we report about
the proper motion measurements and the kinematic properties of the maser distribution. In Sect.\,\ref{discussion}, we model the full-space
velocity field of the CH$_3$OH gas, and discuss the implications for the disk-like scenario in \cep. Conclusions are drawn in Sect.\,\ref{concl}.

%_____________________________________________________________
%                                              FIGURES  N.1
%-----------------------------------------------------------------------------------------------------------
\begin{figure}
\centering
\includegraphics [angle= 0, scale= 0.4]{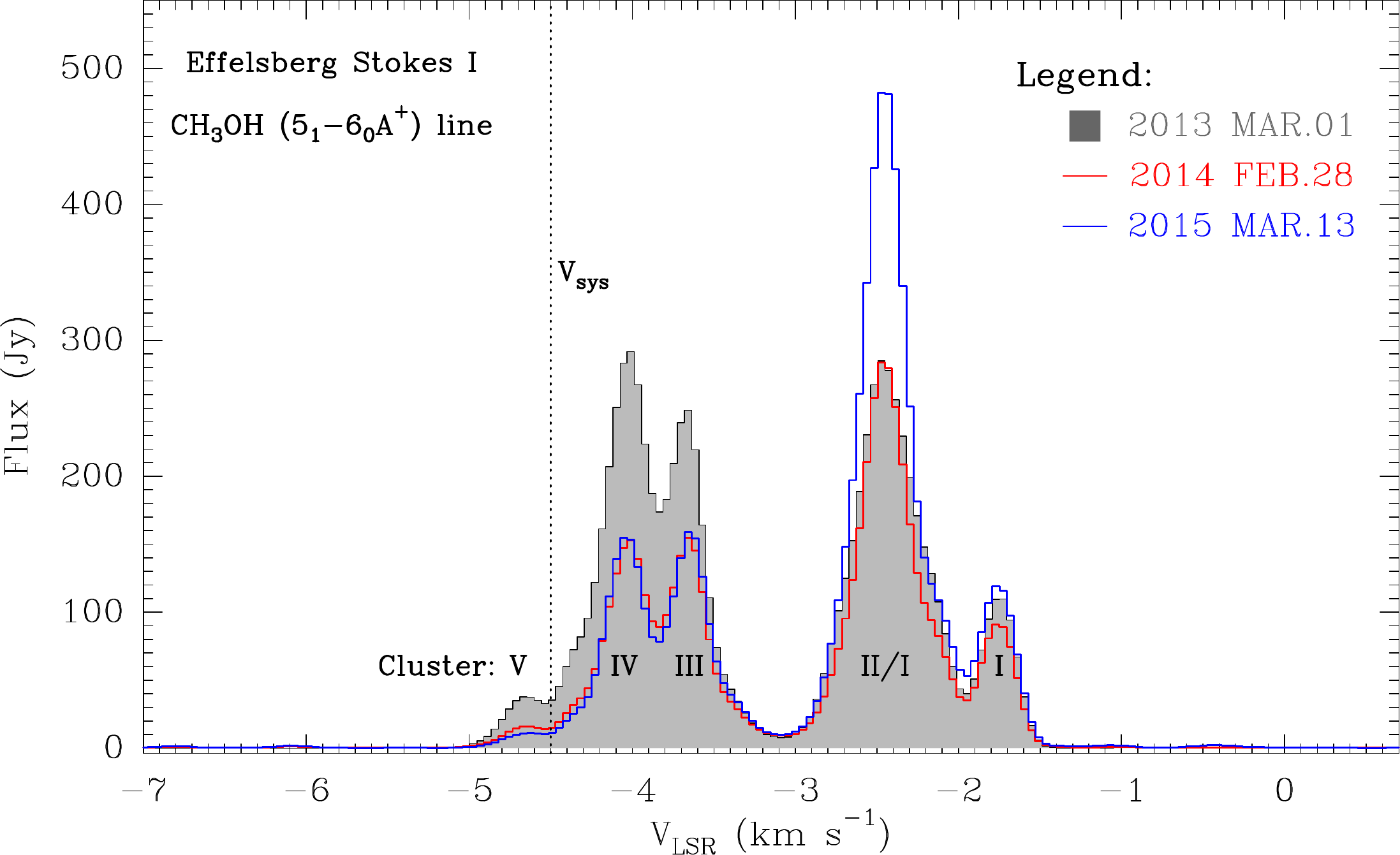}
\caption{Effelsberg total-power spectra towards \cep\, obtained from the EVN observations at C-band. 
Observing dates are indicated on the top right corner. The dotted vertical line indicates the rest velocity 
(V$_{sys}$) of the HW2 object. According to Fig.\,\ref{fig2}, maser emission associated with the different
clusters is also indicated.}
\label{fig1}
\end{figure}
%_____________________________________________________________
%-----------------------------------------------------------------------------------------------------------

\section{Observations and calibration}\label{obs}

We performed multi-epoch, phase-referencing, EVN observations of the $5_1-6_0$\,A$^+$ CH$_3$OH maser transition towards \cep, 
and correlated all 4 polarization combinations (RR, LL, RL, LR). At each epoch, we observed three calibrators together with the maser target.
First, the reference polarization calibrator, J\,1331$+$3030, was observed at the beginning of the experiment to register the systematic
rotation of the linear polarization angle ($\chi_{\rm pol}$) in the EVN dataset. Second, J\,2202$+$4216 was observed as a fringe finder
calibrator every hour to correct both the instrumental phase delay and the polarization leakage. Third, we observed the reference position
calibrator J\,2302$+$6405, which is in the International Celestial Reference Frame (ICRF) catalog, and is offset by $2\degr$ from 
Cepheus\,A. Scans on the maser target and the position calibrator were alternated over a cycle of 5\,min. The EVN data were processed with
the SFXC software correlator \citep{Keimpema2015} at the Joint Institute for VLBI in Europe by using an averaging time of 2\,s and two
frequency setups. A high spectral sampling (0.98\,kHz) over a narrow band (2\,MHz) was used to accurately sample the maser linewidth.
A course spectral sampling  (15.6\,kHz) over a wide band of 16\,MHz was used to improve on the continuum sensitivity of the calibrators
maps. Observation information is summarized in Table\,\ref{tab1}.

Data were reduced with the NRAO Astronomical Image Processing System (AIPS) following standard procedures. Fringe fitting and self calibration
were performed on a strong maser channel at an LSR velocity of --1.73\,\kms\, at each epoch. We produced total intensity (Stokes\,$I$) maps of
the maser emission to image an area of radius $1\farcs5$ around HW2. At each epoch, we set a restoring beam size of 4.5\,mas  (round), equal
to the geometrical average of the \emph{clean} beam size obtained with ROBUST 0 weighting (task IMAGR of AIPS). Figure\,\ref{fig1} shows the
Effelsberg spectra of the CH$_3$OH maser emission towards \cep\, at each epoch.

We postpone the analysis of the polarization information to a subsequent paper. Recently, \cite{Lankhaar2016} presented a theoretical derivation
of the hyperfine structure of the methanol molecule. They show that a number of hyperfine transitions, close in frequency, might contribute 
to the 6.7 GHz CH$_{3}$OH maser emission. In order to account for this multiplicity, a new pumping model for the 6.7\,GHz CH$_{3}$OH masers
is required. These findings might have an influence on the modeling of the polarized maser emission \citep{Vlemmings2010} which is under
review (Lankhaar et al.\,2017, in prep.; Vlemmings et al.\,2018, in prep.).

%_____________________________________________________________
%                                              FIGURES  N.2
%-----------------------------------------------------------------------------------------------------------
%\onlfig{
\begin{figure*}
%\sidecaption
\centering
\includegraphics [angle= 0, scale=0.5]{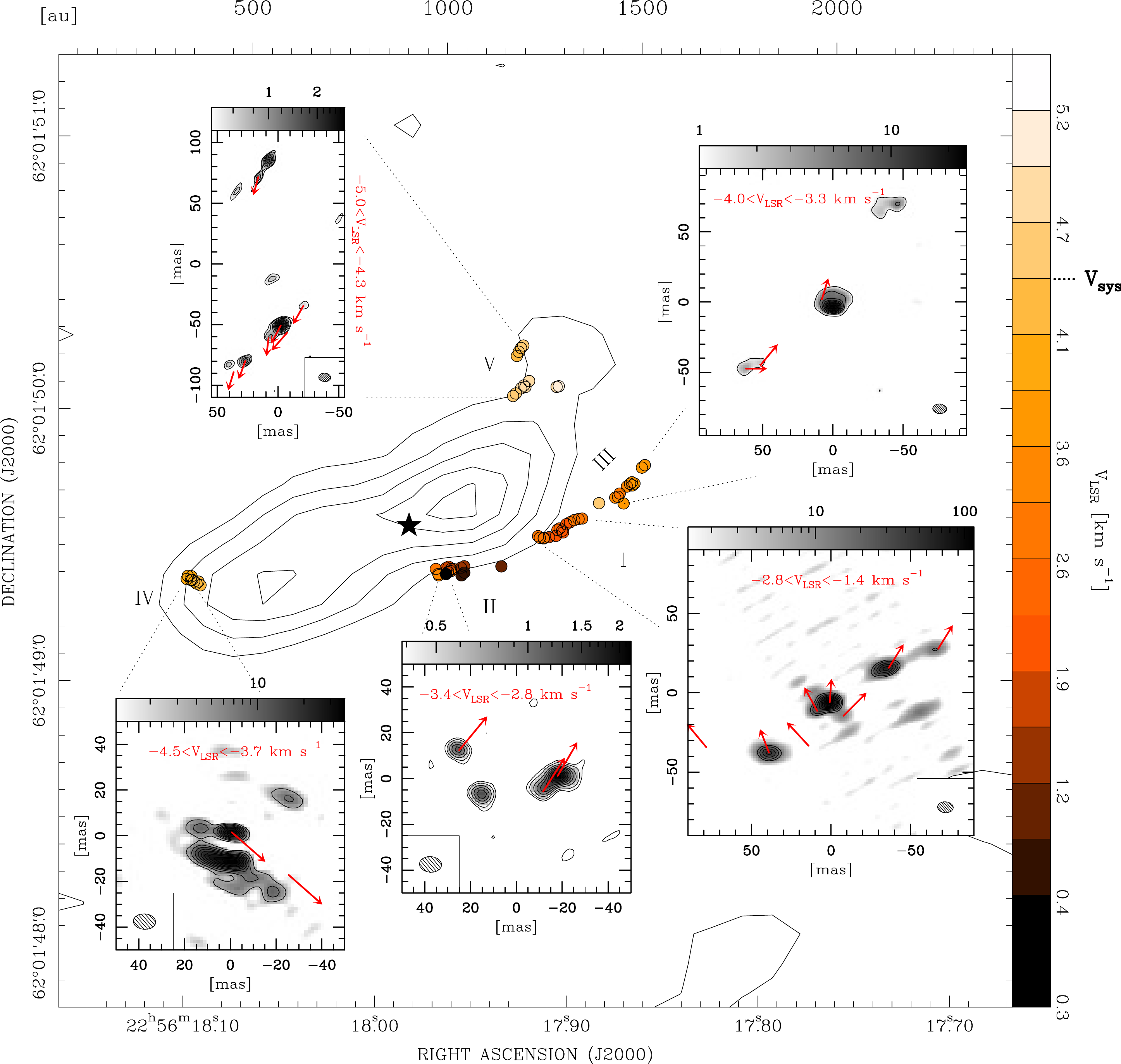}
\caption{Distribution of the 6.7\,GHz CH$_3$OH maser cloudlets detected within $\sim$1\,arcsec of \cep. 
The main panel shows  the distribution of the maser cloudlets (colored dots) superimposed on a HC$_3$N\,(27--26)\,$v_7$\,=\,$1_e$
map (contours) integrated in the velocity range from $-7$ to $-3$\,\kms\, (from \citealt{JimenezSerra2009}). 
HC$_3$N contours start at 2\,$\sigma$ by 2\,$\sigma$ steps. The upper axis gives the linear scale of the map at
a distance of 700\,pc. Maser colors indicate the gas LSR velocity locally, according to the righthand wedge. Maser 
groups are labeled according to  \citet{Sugiyama2008a}. The systemic velocity (V$_{sys}$) is set at $-4.5$\,\kms. The black
star marks the HW2 position according to \citet{Curiel2006}. 
\textbf{Insets:} blowup of the local maser emission (beam size at a bottom corner) summed within the V$_{\rm LSR}$ range specified 
in each box (see Sect.\ref{results}). The grey logarithmic scale (top wedge) gives the brightness scale from  10\,$\sigma$ up to the peak
emission in Jy\,beam$^{-1}$; the black contours give the brightness levels from 10\% of the  peak emission by 10\% steps. The red
arrows trace the proper motion vectors of the maser cloudlets; the arrow length indicates the position reached after a time of 40\,years
at the average velocity of 1.7\,\kms. Note that a few proper motions are associated with faint maser cloudlets which do not show up
in the summed maps.}
\label{fig2}
\end{figure*}
%}
%_____________________________________________________________
%-----------------------------------------------------------------------------------------------------------

%\section{Results}\label{results}

\section{Kinematic properties of the maser distribution}\label{results}

We detected  64 individual maser emission centers, or \textit{cloudlets}\footnote{Hereafter, we make use of the following convention:
\begin{itemize}
\setlength\itemsep{0.1em}
\item a maser \emph{spot} is a compact emission, at milli-arcsecond scale, detected on a single channel map and
best-fitted with an elliptical Gaussian brightness distribution;
\item a maser \emph{cloudlet}, elsewhere referred to as ``feature'', is an individual gas condensation which is
composed by a cluster of contiguous spots, both, in space ($<$\,HPBW$/10$, typically), and LSR velocity ($\la$\,FWHM$/5$,
based on an average linewidth of 0.2--0.3\,\kms).
\end{itemize}},  
above a threshold of 7\,$\sigma$. For strong maser channels, this threshold is typically set by the limited dynamical range of the images.
In Table\,\ref{tab2}, we list the relative position of each maser cloudlet with respect to the peak of the reference maser channel (used to
calibrate the visibilities). The absolute position of this maser spot is determined with an accuracy of $\pm1$\,mas and is given in Table\,\ref{tab1}.
The centroid position of each CH$_3$OH cloudlet is determined by an intensity-weighted average of the spots' distribution within a beam,
following \citet{Sanna2010a}. For cloudlets brighter than 1\,Jy\,beam$^{-1}$,  positional uncertainties are typically better than 0.1\,mas.
%We selected the brightest and stable velocity channels of each maser feature to provide the most accurate centroid of each cloudlet.

Maser cloudlets are spread within a projected radius of 1000\,au from the HW2 object and are arranged in five clusters by position (Fig.\,\ref{fig2}).
These  clusters have been labeled, from I to V,  after \citet{Sugiyama2008b}. The clusters~I and~II, which project closer to the HW2 object, show the 
most red-shifted emission and span a range of LSR velocities between  0 and $-3.5$\,\kms. The three clusters, III,~IV, and~V, projected
further away from HW2, are centered at LSR velocities (intensity-weighted) of $-3.63$, $-4.05$, and $-4.54$\,\kms, respectively, and each 
one emits over a narrow range of velocities (ca. 0.7\,\kms). The five insets of Fig.\,\ref{fig2} show the details  of the maser emission, 
summed over the relevant LSR velocity ranges, at the position of each cluster (from the first epoch data). These maps were produced after a boxcar
smoothing of three velocity channels of the initial dataset, and by imaging the new dataset with an $u$--$v$ tapering of 15\,M$\lambda$.
At variance with the integrated maser flux, which shows significant variations among the epochs (up to a factor of 2), the relative positions among
the clusters, and their ranges of LSR velocities, have remained constant during the last 10 years, since the first high resolution observations
by \citet[their Fig.\,1]{Torstensson2011} on November 2004. 

In order to provide proper motion measurements accurate to a few 0.1\,\kms, we studied the spatial and spectral distribution of the maser spots
within each cloudlet detected at the three epochs (e.g., \citealt{Sanna2010a}, their Fig.\,6). 
At a distance of 700\,pc, a proper motion of 1\,\kms\, corresponds to a displacement of 0.6\,mas every 2 years. We do not report proper 
motions for cloudlets which changed their internal spots' distribution over the epochs so that their centroid position shifted by more than 0.3\,mas. 
Given that we expect proper motions of a few \kms, based on the LSR velocity dispersion of the maser emission, this criterion allows us to exclude 
proper motions severely affected by the internal cloudlet structure.
We selected a sub-sample of 24 maser cloudlets that satisfy this condition and calculated their \emph{relative} proper motion vectors, with respect
to the reference cloudlet at zero offset (num.\,2), by interpolating their positions among the epochs with a linear regression fit (e.g., 
\citealt{Moscadelli2010}, their Fig.\,3). Proper motions were referred to the reference system of the gas at rest by subtracting
the average velocity vector of the whole sample. In columns~6 and~7 of  Table\,\ref{tab2},
we list the eastward and northward velocity components of the corrected proper motions. Their magnitude does not exceed 5\,\kms\, and has an
average value of 1.7\,\kms. The mean uncertainty  per velocity component is 0.2\,\kms. In each inset of Fig.\,\ref{fig2},  we plot the direction of
the proper motion vectors at each measured position (red arrows). The proper motion magnitude gives the position that would be reached by a maser cloudlet,
moving at the average velocity of 1.7\,\kms, after a time of 40\,years. 

%A proper motion of 1 mas\,yr$^{-1}$ corresponds to 3.3\,\kms\, at a distance of 700\,pc. 

%_____________________________________________________________
%                                              FIGURES  N.3
%-----------------------------------------------------------------------------------------------------------
%\onlfig{
\begin{figure}
\centering
\includegraphics [angle= 0, scale=0.32]{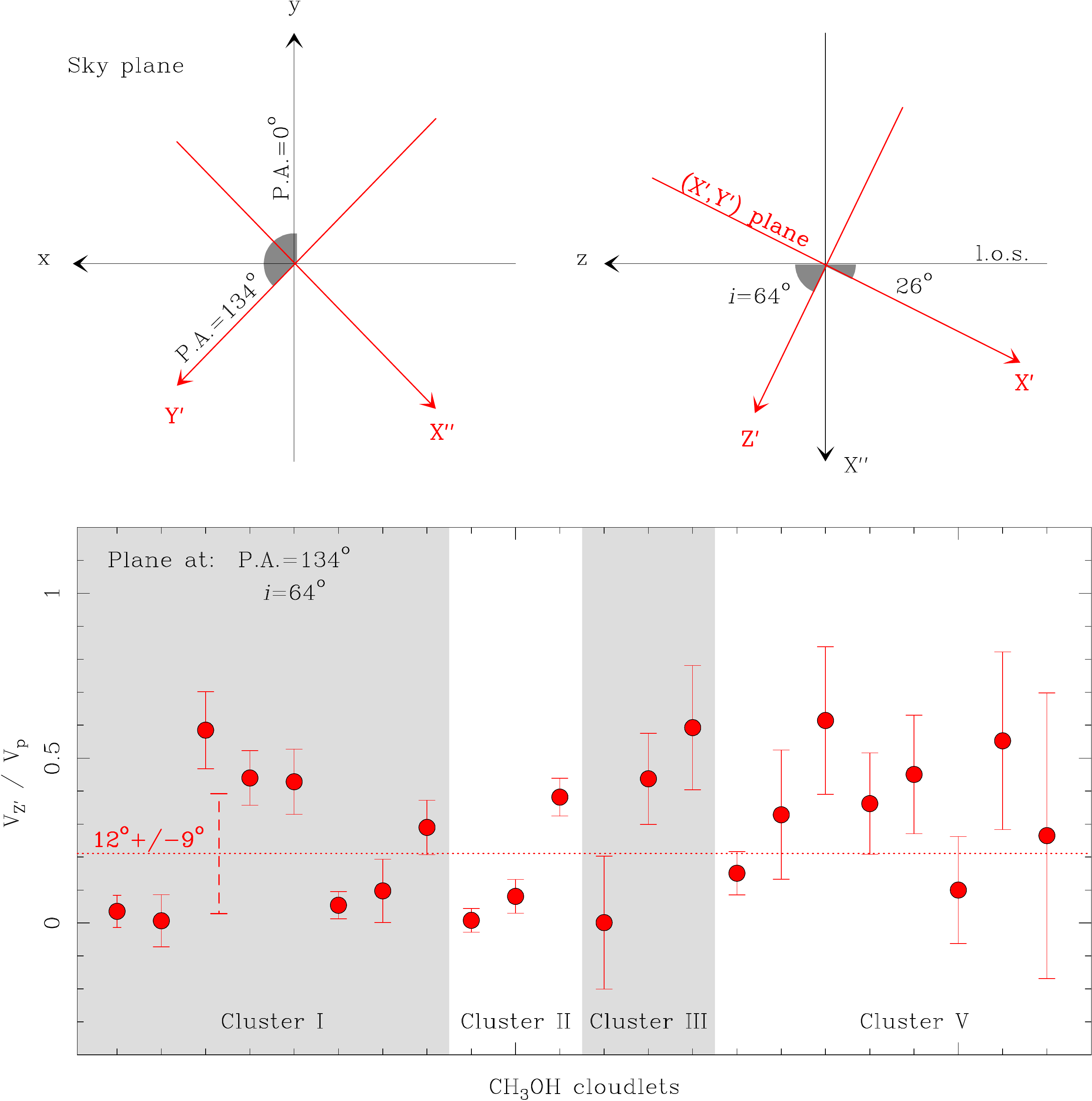}
\caption{Orientation of the best plane ($X^{'},Y^{'}$) which minimizes the velocity component ($V_{\rm Z^{'}}$) of the CH$_3$OH masers perpendicular
to the plane.
\textbf{Upper:} Left and right plots show a view of the best plane as projected on the plane of the sky and with respect to the line-of-sight, respectively.
\textbf{Lower:} ratio of the velocity components perpendicular ($V_{\rm Z^{'}}$) and parallel ($V_{\rm p}$) to the best plane, plotted
for each maser cloudlet (grouped by clusters). Errorbars give the $1\,\sigma$ uncertainty of the ratio $V_{\rm Z^{'}}$\,$/$\,$V_{\rm p}$.
The dotted line marks the weighted average of $V_{\rm Z^{'}}$\,$/$\,$V_{\rm p}$ (with its uncertainty), which corresponds to an average
deviation of $12\degr$ away from the best plane.}
\label{fig3}
\end{figure}
%}
%_____________________________________________________________
%-----------------------------------------------------------------------------------------------------------

\section{Discussion}\label{discussion}

\subsection{Preferential plane of gas motion}\label{plane}

Because CH$_3$OH molecules are (mainly) formed on grain surfaces and are evaporated in the gas phase at a temperature (100--300\,K; \citealt{Herbst2009})
well below the dust sublimation temperature ($\sim$1500\,K), we do expect that the presence of CH$_3$OH gas should be coupled
with the presence of dust condensation. The spatial distribution of dust continuum emission and high-density molecular lines towards \cep\,
has been interpreted as the silhouette of an accretion disk with a radius of several 100\,au. As a comparison, in Fig.\,\ref{fig2} we
overlap the distribution of the CH$_3$OH maser cloudlets with an integrated velocity map of the HC$_3$N emission in the $v_7$\,=\,1
excited state (from \citealt{JimenezSerra2009}). The CH$_3$OH maser transition at 6.7\,GHz is inverted by IR radiation, in the range 20--30\,$\mu$m, 
which is enhanced in the dusty environment heated by a young massive star (e.g., \citealt{Cragg2005}). Similarly, the HC$_3$N vibrationally
excited states are populated by radiative pumping due to an IR field at wavelengths of a few 10\,$\mu$m \citep{Goldsmith1982}. 
Figure\,\ref{fig2} emphasizes the spatial correlation between the dense molecular envelope surrounding HW2 and the loci of CH$_3$OH
maser emission.

Having on hand the information of the full-space motion of the CH$_3$OH gas, we searched for the existence of a preferential plane
of gas motion, as it would be expected in the presence of a disk. With reference to the upper panel of Fig.\,\ref{fig3}, we made two
consecutive rotations of the original reference system in equatorial coordinates (R.A.$\equiv$\,x,\,Dec.$\equiv$\,y):
1) we projected the maser proper motions on the ($X^{''},Y^{'}$) reference system, rotated by an angle ``P.A.'' on the plane of the sky;
2) taking into account also the LSR velocity components, we followed a second projection onto the reference system ($X^{'},Y^{'},Z^{'}$),
obtained by rotating the coordinates ($X^{''},Z$) around the axis $Y^{'}$ by an angle ``\emph{i}''.
We iteratively explored the full range of ($X^{'},Y^{'}$) plane orientations varying ``P.A.'' and ``\emph{i}'' in the ranges 0--360$\degr$
and 0--90$\degr$, respectively. At each run, we calculated the ratio between the maser velocity components perpendicular ($V_{\rm Z^{'}}$)
and parallel ($V_{\rm p}$) to the plane ($X^{'},Y^{'}$), in order to estimate how well the velocity vectors accommodate into the ($X^{'},Y^{'}$)
plane. The upper panel of Fig.\,\ref{fig3} shows the orientation of the best plane (in red) which minimizes the weighted average ratio
$V_{\rm Z^{'}}$\,$/$\,$V_{\rm p}$ for the maser sample of 24 cloudlets.
This plane intersects the plane of the sky along a line at P.A. of 134$\degr$ and it is inclined by 26$\degr$ with respect to the line-of-sight.  
In the lower panel of Fig.\,\ref{fig3}, we plot the ratio $V_{\rm Z^{'}}$\,$/$\,$V_{\rm p}$\,$\pm$\,1$\sigma$ for each maser cloudlet
separately (dots and errorbars). We do not plot the two masers belonging to Cluster\,IV, because they have motions of less than 0.5\,\kms\,
and high relative uncertainties. The weighted average of the sample is drawn with a dashed horizontal line in the lower panel of
Fig.\,\ref{fig3}, and corresponds to a residual inclination of 12$\degr$\,$\pm$\,9$\degr$ with respect to the plane ($X^{'},Y^{'}$).
Alternatively,  this inclination corresponds to an average velocity component of 0.5\,\kms\, away from the plane.

The best-fit plane (hereafter, the equatorial plane), over which CH$_3$OH masers move, is coincident with the orientation of the flattened core
observed by \citet{Patel2005} within about $10\degr$ (``P.A.'' and  ``\emph{i}'' of 121$\degr$ and 62$\degr$, respectively). 
It is also worth noting that the orientation of the radio jet, at a P.A. near $45\degr$ \citep[their Table\,2]{Curiel2006}, 
coincides with the minor axis of the equatorial plane as projected onto the sky ($X^{''}$ in Fig.\,\ref{fig3}).  
\emph{These findings support the interpretation that the core emission imaged by \citeauthor{Patel2005} belongs to a continuous structure, which 
extends along the equatorial plane of gas motion.} In the next section, we analyze the gas motion in the equatorial plane and show that
maser cloudlets participate of a global infall towards HW2.

%_____________________________________________________________
%                                              FIGURES  N.4
%-----------------------------------------------------------------------------------------------------------
%\onlfig{
\begin{figure*}
\sidecaption
\centering
\includegraphics [angle= 0, scale=0.55]{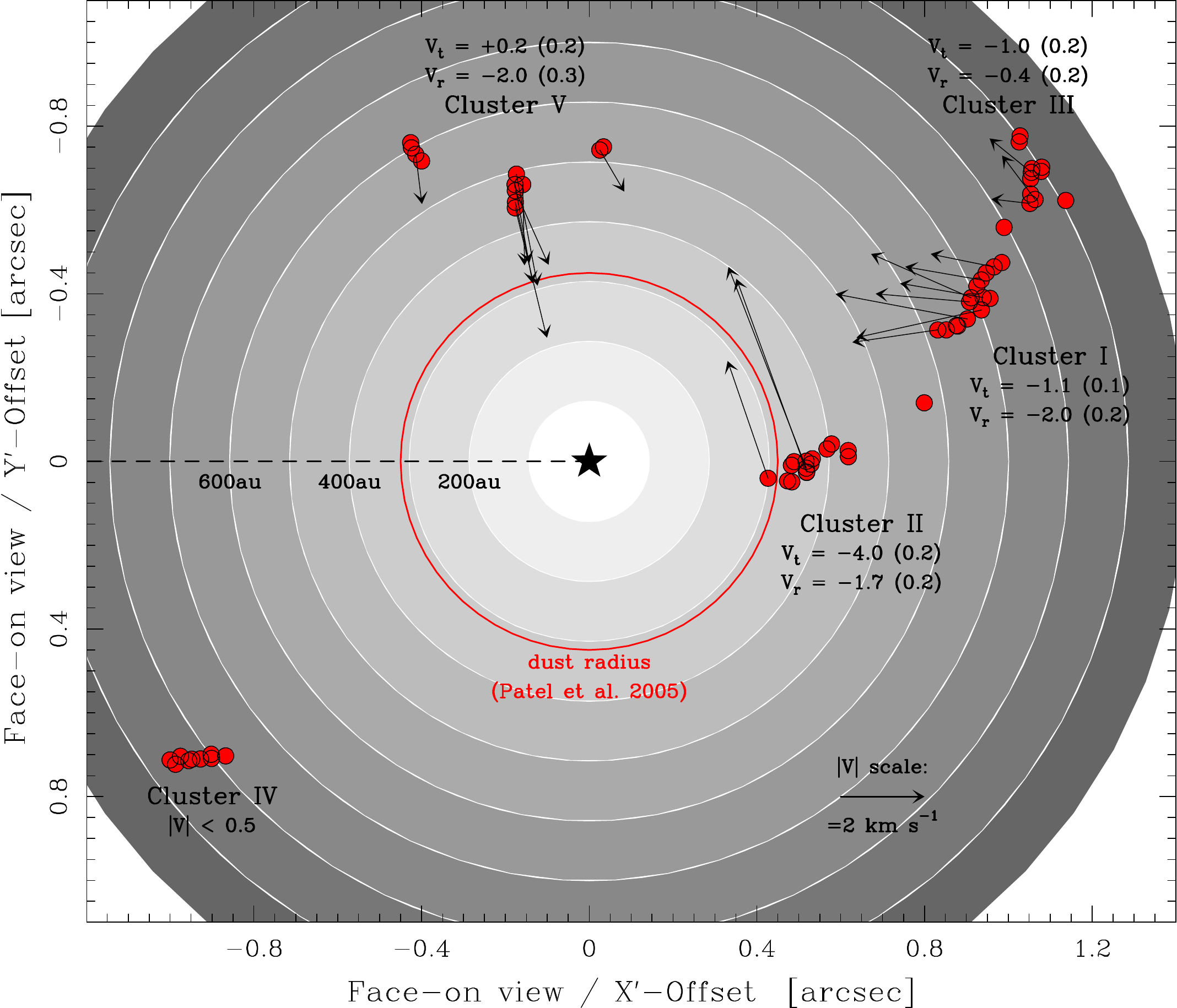}
\caption{Face-on view of the plane at a P.A. of $134\degr$ and inclination of 64$\degr$, which contains 
the full-space motions of the 6.7\,GHz CH$_3$OH maser cloudlets  (see Fig.\,\ref{fig3}). The star
at the center of the plot pinpoints the position of the HW2 object which powers the radio continuum jet
perpendicular to the plane \citep{Curiel2006}.
Grey circles mark different distances from the YSO position along this plane. Red dots mark the different locations
of the CH$_3$OH maser cloudlets, and the black arrows trace their motions in the plane. Each maser cluster 
is labeled according to Fig.\,\ref{fig2}. The velocity scale is reported at the bottom right corner. For each cluster,
we report the radial ($V_{\rm r}$) and tangential ($V_{\rm t}$) group velocities with respect to the YSO position
($1\,\sigma$ uncertainty in brackets). The red circle, at a radius of about 300\,au, marks the deconvolved size of the 
dust emission at 0.9\,mm observed by \citet{Patel2005}.}
\label{fig4}
\end{figure*}
%}
%_____________________________________________________________
%-----------------------------------------------------------------------------------------------------------

\subsection{Analysis of the planar motion around HW2}

Knowing the orientation of the equatorial plane, we can produce a face-on view of the CH$_3$OH maser velocities in 
this plane. In Fig.\,\ref{fig4}, we plot the position of the CH$_3$OH cloudlets (red dots) projected onto the equatorial plane.
We set the origin of the plot at the position of HW2 as defined by \citet{Curiel2006}. Each cluster of maser emission is
labeled according to Fig.\,\ref{fig2} for comparison. The result shown in Fig.\,\ref{fig4} is independent of any assumption
about the maser geometry. Individual centers of maser emission do not origin at a same distance from the star. Maser emission is
excited over a range of radii between 300\,au and 900\,au from HW2. The inner radius coincides with the extent of the dust
emission (hereafter, the dust radius) at a wavelength of 0.9\,mm (red circle in Fig.\,\ref{fig4}). 
A disk radius of about 300\,au also marks the boundary where hydrogen density might exceeds a threshold of 10$^8$\,cm$^{-3}$
\citep[their Fig.\,9]{Kuiper2015}, above which the CH$_3$OH maser transition at 6.7\,GHz is quenched \citep{Cragg2005}.

The black arrows in Fig.\,\ref{fig4} give the magnitude and direction of the velocity vectors in the equatorial plane. Since maser cloudlets belonging 
to a given cluster show ordered velocity vectors, for each cluster we derived a group motion in the radial ($V_{\rm r}$) and tangential
($V_{\rm t}$) directions with respect to HW2. These values are reported together with the cluster labels in Fig.\,\ref{fig4}. A negative radial
velocity corresponds to an inward motion towards HW2; a negative tangential velocity indicates a counterclockwise rotation around HW2.  
For radii between 300\,au and 800\,au (clusters I, II, and V), maser velocities have an average inward component of about $-2$\,\kms\,
within an uncertainty of 0.2\,\kms. \emph{We explicitly note that the existence of a planar infall toward HW2 supports, in turn, the existence
of an accretion disk.} At the loci sampled by the maser emission, the magnitude of the infalling motions appears to be almost constant   
with the distance from the star within the uncertainties. Outside a radius of 800\,au (clusters III and IV), the radial component
is more than four times slower ($<$0.5\,\kms) than in the inner regions. 
On the other hand, there is no clear pattern for the tangential velocity outside a radius of 300\,au from the star: clusters\,I and~III have a
consistent tangential component of about $-1$\,\kms, clusters\,IV and~V have a tangential component near zero, and cluster\,II has the
faster tangential component  of $-4$\,\kms. We do not have enough maser points around the star to study the dependence (if any) of the
tangential velocity with the azimuthal angle.

\subsection{Estimate of the central mass of HW2}

Interestingly, the maser cluster (II) closer to the star exhibits the higher value of tangential velocity.  If we assume that this tangential
component represents an ordered rotational motion around HW2, we can evaluate the enclosed mass within the dust radius. 
Assuming Keplerian rotation, for a rotational velocity of 4\,\kms\, and a radius of 315\,au, derived from the deconvolved size of the dust
continuum map of \citet{Patel2005}, one infers a central mass of 5.5\,M$_{\odot}$. A ZAMS star with a mass of 6\,M$_{\odot}$ and a radius
of 3\,R$_{\odot}$ has a bolometric luminosity of  1\,$\rm\times10^3\,L_{\odot}$  (e.g., \citealt{Schaller1992}), one tenth of the bolometric
luminosity associated  with the HW2 object. An additional contribution to the total luminosity might come from accretion. We make use of
the infalling motion  measured with the masers to estimate the mass infall rate onto the star. In spherical symmetry, for an hydrogen density
of 10$^8$\,cm$^{-3}$ at a distance of 300\,au from the central object, an infall velocity of  2\,\kms\, implies an accretion rate of
3\,$\rm\times10^{-4}\,M_{\odot}$\,yr$^{-1}$. If the same infall rate reaches the star surface, the accretion luminosity
of 9\,$\rm\times10^3\,L_{\odot}$ would dominate the total luminosity. Taking into account that the infalling mass is channeled 
through the disk onto the star within a small solid angle, the accretion rate might be reduced by a factor $\sim$\,0.1 for a flared disk 
with opening angle of $60\degr$. Under these conditions, the total budget of stellar and accretion luminosities would account for a bolometric
luminosity of  about 2\,$\rm\times10^3\,L_{\odot}$.

Different lines of evidence suggest the presence of an ordered magnetic field in the inner 1000\,au of HW2, which might reconcile the kinematic
estimate of the stellar mass with a bolometric luminosity of the order of 1\,$\rm\times10^4\,L_{\odot}$.  Recently,
\citet{FernandezLopez2016} reported about CARMA dust polarization observations at 1.3\,mm in the vicinity of HW2.
The dust polarization vectors align in a direction (57$\degr$\,$\pm$\,6$\degr$) roughly parallel to the minor axis of the accretion disk
on the plane of the sky ($X^{''}$ in Fig.\,\ref{fig3}). If the main cause of polarized emission is grain alignment via magnetic forces,
\citeauthor{FernandezLopez2016} propose a scenario where an uniform magnetic field is threading the accretion disk of HW2. At the scale
of individual pockets of gas of a few au in size, the linearly polarized emission of the OH, H$_2$O, and CH$_3$OH masers still shows coherent
polarization vectors either parallel or perpendicular to the same $X^{''}$ axis of Fig.\,\ref{fig3} \citep{Bartkiewicz2005,Vlemmings2006,Vlemmings2010}.
The presence of a magnetic field might result in rotational velocities less than half the expected Keplerian velocity at a given radius, as
depicted in Fig.\,7 of \cite{Seifried2011}. In the case of HW2, if the tangential velocity measured with the masers were (magnetically)
reduced by a factor as small as $1/4$  of the pure Keplerian rotation, the revised mass estimate would amount to 10\,M$_{\odot}$ and
would imply a ZAMS luminosity of about 1\,$\rm\times10^4\,L_{\odot}$ (e.g.,  Eq.\,13 of \citealt{Hosokawa2010}).

\subsection{Comparison with previous models}

In order to explain the CH$_3$OH maser distribution around HW2, previous models assumed that the maser 
emission arises along a circular ring \citep{Torstensson2008,Vlemmings2010,Torstensson2011,Sugiyama2014}.
Maser positions were fitted by an elliptical ring on the plane of the sky, with free parameters the major (2\,$\times$\,\emph{a})
and minor axes of the ellipse, the position angle of the major axis (P.A.$_{ell}$), and the center of the ellipse. If the ellipse is the
projection of a circular ring tilted with respect to the line-of-sight, the inclination (\emph{i}) of the circular ring is determined from
the ratio of the minor-to-major axes of the ellipse. Since the molecular outflow entrained by the radio jet has the blueshifted lobe to 
the NE, and the free-free optical depth of the radio jet favors the brightest masers to be on the near side of the ring, \citeauthor{Torstensson2011}
first argued that maser clusters~I and~II have to lie on the near side of the ring. Under these conditions, the plane of the circular ring
would be oriented at a position angle (P.A.$_{ell}$) between 99$\degr$ and 110$\degr$, with an inclination (\emph{i}) betwen 68$\degr$
and 73$\degr$; the ring radius (\emph{a}) would range between 650\,au and 680\,au \citep{Vlemmings2010,Torstensson2011,Sugiyama2014}.
The position angle of the equatorial plane derived in Sect.\,\ref{plane} differs by about 30$\degr$ w.r.t. the plane of the circular 
ring; the radius of the circular ring approximates the average distance of the maser cloudlets w.r.t. HW2 (Fig.\,\ref{fig4}).   

To explain the three-dimensional magnetic field configuration around HW2, as inferred from the 6.7\,GHz maser emission, \citeauthor{Vlemmings2010}
also proposed that the maser cloudlets might be detached by a few 100\,au from the equatorial plane. Maser cloudlets would move along the
magnetic field lines at the edge of the disk, where the lines warp towards the central core (Fig.\,6 of \citealt{Vlemmings2010}). This picture
might still hold even if the maser emission does not arise at a constant radius, since the toroidal magnetic field lines would thread the disk
plane at different radii from the central star. The current analysis does not rule out that the maser cloudlets may lie at small heights
(\emph{h}) above the equatorial plane (\emph{h}\,$\la$\,100\,au), provided that the velocity field is still mainly planar (e.g., Fig.\,9 of \citealt{Seifried2011}).  

\citeauthor{Torstensson2011} compared the line-of-sight velocities of the maser emission with those expected for a circular ring either
in Keplerian rotation or with a dominant radial motion. They inferred negligible Keplerian rotation and proposed that the ring itself is
contracting towards HW2 with a velocity of the order of 1\,\kms. \citeauthor{Sugiyama2014} also measured the 6.7\,GHz maser proper
motions with the  Japanese VLBI Network (JVN).  They modeled the three-dimensional velocity field of the CH$_3$OH gas as the composition
of a radial and tangential velocity with respect to the ring, and derived a global infalling motion of 2.0\,$\pm$\,0.6\,\kms\, and a
rotational velocity near zero. These results generally agree with our analysis when the group velocities of each cluster
are averaged all together (Fig.\,\ref{fig4}).  It is worth noting that the EVN observations have a beam size 2 times smaller than the
JVN observations, and 4 times better spectral resolution, which eventually allow us to obtain proper motion measurements about 3
times more accurate than the JVN measurements. On the one hand, since we measure a constant radial motion between radii of 300\,au
and 800\,au, this finding explains why \citeauthor{Sugiyama2014} inferred a similar infalling velocity at constant radius.
On the other hand, by removing the constraint of constant radius, we are able to distinguish different tangential velocities associated 
with the five maser clusters. 

\section{Conclusions}\label{concl}

We tested the existence of an equatorial plane of gas motion around \cep\, by measuring the three-dimensional 
velocity vectors of the 6.7\,GHz CH$_3$OH maser cloudlets. Maser velocities are measured with an accuracy of a few 0.1\,\kms.
Our results can be summarized as follow:  

\begin{enumerate}

\item We show that the motions of the CH$_3$OH maser cloudlets are dominated by a planar component. Its plane is inclined
by 26$\degr$ with the line-of-sight, and intersects the plane of the sky along a line at a P.A. of 134$\degr$. The orientation of
this plane coincides, within about 10$\degr$, with that of the core emission imaged by \citet{Patel2005}, after assuming that
dust and high-density gas move on circular orbits. This result supports the interpretation that the core emission surrounding HW2
represents a continuous structure, rather than being a projection of different hot cores along the line-of-sight.

\item The three-dimensional velocity field of the CH$_3$OH masers shows that the CH$_3$OH gas is contracting towards HW2
at a velocity of about 2\,\kms, between radii of 300 and 900\,au from HW2. This velocity component implies an accretion rate
of the order of $10^{-4}\,M_{\odot}$\,yr$^{-1}$  at a distance of 300\,au from the star. The existence of a planar infall toward
HW2 supports, in turn, the existence of an accretion disk.

\item At a radius of 300\,au from HW2, we measure a rotational component of about 4\,\kms. The simple assumption of pure
Keplerian rotation would result into a central mass of 5--6\,M$_{\odot}$, which would account for only a minor fraction of the
bolometric luminosity of the HW2 object. If the rotational velocities measured with the masers were sub-Keplerian by a factor as
small as $1/4$, as it can be expected under the influence of a weak magnetic field, that could reconcile the values of the 
(kinematic) stellar mass and luminosity. The presence of a magnetic field is supported by the recent detection of an ordered dust
polarization field surrounding HW2, and can be tested with the 6.7\,GHz maser observations once a new model for the polarized 
emission of this maser transition will be available.

\end{enumerate}

\begin{acknowledgements}

The European VLBI Network is a joint facility of independent European, African, Asian, and North American radio astronomy institutes.
Scientific results from data presented in this publication are derived from the following EVN project code(s): ES071. 
Comments from an anonymous referee are gratefully acknowledged.
Financial support by the Deutsche Forschungsgemeinschaft (DFG) Priority Program 1573 is gratefully acknowledged.
A.M. Sobolev is financially supported by the Russian Science Foundation (project no. 15-12-10017).
We thank W.H.T. Vlemmings for fruitful discussions in preparation. We are grateful to I.\,Jim{\'e}nez-Serra for
providing us with the HC$_3$N map shown in Fig.\ref{fig2}. 

\end{acknowledgements}

%---------------------------- REFERENCES ------------------------------

\bibliographystyle{aa}
\bibliography{asanna1303}

%\clearpage

%\begin{appendix}

%\end{appendix}

%\clearpage

%_____________________________________________________________
%                                              FIGURES  N. n
%-----------------------------------------------------------------------------------------------------------
%\begin{figure*}
%\sidecaption
%\centering
%\includegraphics [angle= 0, scale= 0.5]{fig2.pdf}
%\includegraphics [angle= 0, width=\hsize]{fig2.eps}
%\caption{...}
%\label{fig2}
%\end{figure*}

%_____________________________________________________________
%-----------------------------------------------------------------------------------------------------------

%_____________________________________________________________
%                                              FIGURES  N.n
%-----------------------------------------------------------------------------------------------------------
%\begin{figure}
%\centering
%\includegraphics [angle= 0, scale= 1.0]{... .pdf}
%\includegraphics [angle= 0, width=\hsize]{fig3.eps}
%\caption{...}
%\label{fig3}
%\end{figure}
%_____________________________________________________________
%-----------------------------------------------------------------------------------------------------------

\clearpage

%_____________________________________________________________
%                                                    TABLES
%-----------------------------------------------------------------------------------------------------------

\onllongtab{
\addtocounter{table}{0}
\begin{longtable}{r r r r r | r r } 
%\centering
\caption{Parameters of the 6.7\,GHz CH$_3$OH maser cloudlets detected toward Cepheus\,A\,HW2. 
Each maser emission center is labeled by decreasing LSR velocity in Column\,1, and grouped within the five clusters defined
in Fig.\,\ref{fig2}. Columns\,2 and~3 report the LSR velocity and brightness of the brightest spot of each cloudlet at the first
epoch of detection (in brackets). Columns\,4 and~5 give the relative centroid position of each cloudlet, and their uncertainties,
in the east and north directions, respectively. The absolute position of the reference spot, belonging to cloudlet num.\,2,
is reported in Table\,\ref{tab1}. Columns\,6 and~7 give the projected components of the cloudlet proper motion along the east
and north directions, respectively. More details are provided
in Sect.\,\ref{results}.}\label{tab2} \\
\hline 
\hline 
\multicolumn{1}{c}{Feature} & \multicolumn{1}{c}{V$_{\rm LSR}$} & 
\multicolumn{1}{c}{F$_{\rm peak}$} & \multicolumn{1}{c}{$\Delta \rm x$} & \multicolumn{1}{c}{$\Delta \rm y$} & 
\multicolumn{1}{c}{V$_{\rm x}$} & \multicolumn{1}{c}{V$_{\rm y}$} \\ 
\multicolumn{1}{c}{\#} & \multicolumn{1}{c}{(km\,s$^{-1}$)} & 
\multicolumn{1}{c}{(Jy\,beam$^{-1}$)} & \multicolumn{1}{c}{(mas)} & \multicolumn{1}{c}{(mas)} & 
\multicolumn{1}{c}{(km\,s$^{-1}$)} & \multicolumn{1}{c}{(km\,s$^{-1}$)}  \\ 
\hline 
\endfirsthead
\caption{continued.}\\
 \hline 
 \hline 
\multicolumn{1}{c}{Feature} & \multicolumn{1}{c}{V$_{\rm LSR}$} & 
\multicolumn{1}{c}{F$_{\rm peak}$} & \multicolumn{1}{c}{$\Delta \rm x$} & \multicolumn{1}{c}{$\Delta \rm y$} & 
\multicolumn{1}{c}{V$_{\rm x}$} & \multicolumn{1}{c}{V$_{\rm y}$} \\ 
\multicolumn{1}{c}{\#} & \multicolumn{1}{c}{(km\,s$^{-1}$)} & 
\multicolumn{1}{c}{(Jy\,beam$^{-1}$)} & \multicolumn{1}{c}{(mas)} & \multicolumn{1}{c}{(mas)} & 
\multicolumn{1}{c}{(km\,s$^{-1}$)} & \multicolumn{1}{c}{(km\,s$^{-1}$)}  \\ 
\hline
\endhead
\hline
\endfoot
\hline
\endlastfoot

 & & & & & & \\ 

 \multicolumn{4}{l}{\textbf{Cluster\,I:  mean LSR velocity of --2.31\,km\,s$^{-1}$}} & & \\
 1  & --1.68 & 1.04  \small{\textbf{(3$^{rd}$)}} & $   19.38\pm0.17$ & $   -4.85\pm0.10$ & ... & ...   \\       
 2  & --1.73   & 22.84  \small{\textbf{(1$^{st}$)}} & $   -0.00\pm0.01$   & $   -0.01\pm0.01$  & $  0.61\pm0.06$ & $  1.49\pm0.06$   \\ 
      &    &    &                    ...         &           ...                  &           ...              &           ...                \\ 
      &    &    &                    ...         &           ...                  &           ...              &           ...                 \\ 
 3  & --1.90 & 0.57  \small{\textbf{(1$^{st}$)}} & $  -25.36\pm0.18$ & $    4.35\pm0.09$ & $  1.53\pm0.37$ & $  0.64\pm0.18$   \\
 4  & --2.03  & 1.81  \small{\textbf{(1$^{st}$)}} & $  -51.33\pm0.06$ & $   17.08\pm0.05$ & ... & ...  \\ 
 5  & --2.17 &   4.07  \small{\textbf{(1$^{st}$)}} & $  -47.75\pm0.06$ & $   23.57\pm0.05$ & $ -0.40\pm0.20$ & $  0.08\pm0.16$   \\  
 6  &  --2.30 & 52.18  \small{\textbf{(1$^{st}$)}} & $  -38.97\pm0.01$ & $   32.19\pm0.01$ & $ -0.10\pm0.07$ & $  1.40\pm0.07$  \\ 
      &    &    &                    ...         &           ...                  &           ...              &           ...               \\ 
      &    &    &                    ...         &           ...                  &           ...              &           ...                 \\ 
 7  & --2.30  & 3.09  \small{\textbf{(1$^{st}$)}} & $  -63.16\pm0.15$ & $   47.39\pm0.12$ & ... & ...   \\    
 8  & --2.47 &  32.57 \small{\textbf{(1$^{st}$)}} & $  -76.76\pm0.02$ & $   54.28\pm0.02$ & $ -0.20\pm0.08$ & $  0.25\pm0.07$   \\
      &    &    &                    ...         &           ...                  &           ...              &           ...                 \\
      &    &    &                    ...         &           ...                  &           ...              &           ...                 \\        
  9  &  --2.47 &  24.08  \small{\textbf{(1$^{st}$)}} & $  -30.86\pm0.02$ & $   26.70\pm0.02$ & $  0.53\pm0.15$ & $  0.83\pm0.13$    \\
      &    &    &                    ...         &           ...                  &           ...              &           ...               \\
 10  & --2.65  & 6.43  \small{\textbf{(1$^{st}$)}} & $  -92.35\pm0.06$ & $   62.36\pm0.04$ &           ...              &           ...                 \\
      &    &    &                    ...         &           ...                  &           ...              &           ...               \\
 11  & --2.82 &  7.65 \small{\textbf{(1$^{st}$)}} & $ -108.34\pm0.02$ & $   66.52\pm0.01$ & $ -0.19\pm0.08$ & $  0.24\pm0.07$  \\ 
      &    &    &                    ...         &           ...                  &           ...              &           ...                \\
      &    &    &                    ...         &           ...                  &           ...              &           ...                \\
12  & --2.91 & 0.94  \small{\textbf{(1$^{st}$)}} & $   40.11\pm0.04$ & $    3.54\pm0.04$ & $  1.18\pm0.13$ & $  0.79\pm0.09$   \\ 
13  & --2.91 & 0.20  \small{\textbf{(3$^{rd}$)}} & $   33.83\pm0.27$ & $   -2.56\pm0.17$ & ... & ...  \\  
14  & --3.00 &  0.25  \small{\textbf{(1$^{st}$)}} & $ -121.31\pm0.10$ & $   67.99\pm0.10$ & ... & ...   \\ 
15  & --3.00 & 0.15  \small{\textbf{(3$^{rd}$)}} & $   19.37\pm0.18$ & $   -3.01\pm0.10$ & ... & ...   \\

\multicolumn{4}{l}{\textbf{Cluster\,II:  mean LSR velocity of --2.44\,km\,s$^{-1}$}} & &  \\
16  & --0.10  & 0.08  \small{\textbf{(3$^{rd}$)}} & $  378.41\pm0.16$ & $ -135.89\pm0.13$ & ... & ...  \\
17  & --0.45  & 0.78  \small{\textbf{(3$^{rd}$)}} & $  311.09\pm0.02$ & $ -130.28\pm0.03$ & ... & ...   \\ 
18  & --0.45  &  0.08  \small{\textbf{(3$^{rd}$)}} & $  321.51\pm0.19$ & $ -140.62\pm0.26$ & ... & ...   \\
19  & --0.94  &  0.13  \small{\textbf{(1$^{st}$)}} & $  174.92\pm0.14$ & $ -107.37\pm0.13$ & ... & ...   \\ 
20  & --1.20  &  0.21  \small{\textbf{(3$^{rd}$)}} & $  323.52\pm0.07$ & $ -111.97\pm0.09$ & ... & ...   \\ 
21  & --1.33  & 0.07  \small{\textbf{(3$^{rd}$)}} & $  311.68\pm0.18$ & $ -107.07\pm0.17$ & ... & ...   \\
22  & --1.81  & 1.67  \small{\textbf{(3$^{rd}$)}} & $  368.78\pm0.08$ & $ -108.69\pm0.07$ & ... & ...   \\
23  & --2.17  & 1.65  \small{\textbf{(3$^{rd}$)}} & $  376.92\pm0.07$ & $ -112.75\pm0.08$ & ... & ...   \\
24  & --2.42  &  9.42  \small{\textbf{(3$^{rd}$)}} & $  359.55\pm0.09$ & $ -116.89\pm0.09$ & ... & ...   \\ 
25  & --2.61  &  7.59  \small{\textbf{(3$^{rd}$)}} & $  350.78\pm0.05$ & $ -117.32\pm0.04$ & ... & ...   \\ 
26  & --2.69  & 1.67  \small{\textbf{(3$^{rd}$)}} & $  360.80\pm0.13$ & $ -125.16\pm0.11$ & ... & ...   \\
27  & --2.78  &  1.25  \small{\textbf{(1$^{st}$)}} & $  371.31\pm0.10$ & $ -128.49\pm0.10$ & $ -2.89\pm0.15$ & $  3.93\pm0.16$   \\ 
28  & --2.82  &  0.86  \small{\textbf{(1$^{st}$)}} & $  377.87\pm0.11$ & $ -135.48\pm0.10$ & $ -3.03\pm0.25$ & $  3.50\pm0.22$  \\ 
29  & --3.18  &  0.78  \small{\textbf{(1$^{st}$)}} & $  415.83\pm0.03$ & $ -117.00\pm0.03$ & $ -2.41\pm0.11$ & $  1.54\pm0.11$   \\ 
30  & --3.22  &  0.68  \small{\textbf{(1$^{st}$)}} & $  406.50\pm0.04$ & $ -135.77\pm0.04$ & ... & ...  \\
31  & --3.44  &  0.59  \small{\textbf{(1$^{st}$)}} & $  405.01\pm0.09$ & $ -140.44\pm0.10$ & ... & ...   \\

\multicolumn{4}{l}{\textbf{Cluster\,III:  mean LSR velocity of --3.63\,km\,s$^{-1}$}} & &  \\
32  & --3.35  &   0.41  \small{\textbf{(1$^{st}$)}}  & $ -259.51\pm0.14$ & $  160.15\pm0.20$ & ... & ...   \\
33  & --3.44  &   2.37  \small{\textbf{(1$^{st}$)}} & $ -242.84\pm0.03$ & $  145.19\pm0.03$ & $ -0.15\pm0.08$ & $  0.00\pm0.09$   \\  
34  & --3.52  &  0.37  \small{\textbf{(3$^{rd}$)}} & $ -286.04\pm0.22$ & $  185.77\pm0.25$ & ... & ...   \\
35  & --3.53  &   2.52  \small{\textbf{(1$^{st}$)}} & $ -253.31\pm0.05$ & $  147.95\pm0.06$ & $ -0.87\pm0.10$ & $  0.61\pm0.13$   \\ 
36  & --3.61  &  0.83  \small{\textbf{(3$^{rd}$)}} & $ -273.56\pm0.21$ & $  123.62\pm0.19$ & ... & ...  \\ 
37  & --3.62  &   1.49  \small{\textbf{(2$^{nd}$)}} & $ -340.85\pm0.09$ & $  255.21\pm0.12$ & ... & ...   \\ 
38  & --3.66  & 10.84  \small{\textbf{(1$^{st}$)}} & $ -305.94\pm0.03$ & $  189.99\pm0.02$ &          ...               &            ...              \\ 
39  & --3.66  &   5.38  \small{\textbf{(1$^{st}$)}} & $ -351.66\pm0.04$ & $  264.29\pm0.06$ &          ...               &            ...              \\
40  & --3.66  &   3.02  \small{\textbf{(1$^{st}$)}} & $ -297.34\pm0.09$ & $  195.69\pm0.12$ & $ -0.28\pm0.23$ & $  0.87\pm0.27$ \\
41  & --3.66  &   2.60  \small{\textbf{(1$^{st}$)}} & $ -304.12\pm0.11$ & $  201.49\pm0.09$ & ... & ...   \\ 
42  & --3.70  &   1.45  \small{\textbf{(1$^{st}$)}} & $ -313.51\pm0.20$ & $  196.58\pm0.22$ & ... & ...   \\ 
43  & --4.49  &   0.10  \small{\textbf{(2$^{nd}$)}} & $ -183.38\pm0.30$ & $  124.61\pm0.19$ & ... & ...   \\

 \multicolumn{4}{l}{\textbf{Cluster\,IV:  mean LSR velocity of --4.05\,km\,s$^{-1}$}} & &  \\ 
 44  & --3.88  &   5.88  \small{\textbf{(1$^{st}$)}} & $ 1326.16\pm0.04$ & $ -142.10\pm0.03$ & ... & ...   \\
 45  & --3.88  &   3.42  \small{\textbf{(1$^{st}$)}} & $ 1329.83\pm0.13$ & $ -153.46\pm0.09$ & ... & ...   \\ 
 46  & --3.92  &   5.28  \small{\textbf{(1$^{st}$)}} & $ 1294.00\pm0.06$ & $ -170.31\pm0.06$ & ... & ...   \\ 
 47  & --4.01  & 19.93  \small{\textbf{(1$^{st}$)}} & $ 1312.58\pm0.03$ & $ -143.87\pm0.02$ & $ -0.44\pm0.09$ & $ -0.03\pm0.08$ \\ 
 48  & --4.01  &   5.28  \small{\textbf{(1$^{st}$)}} & $ 1302.81\pm0.09$ & $ -163.01\pm0.07$ & ... & ...   \\ 
 49  & --4.05  & 13.62  \small{\textbf{(1$^{st}$)}} & $ 1314.18\pm0.06$ & $ -157.25\pm0.03$ & ... & ...   \\  
 50  & --4.19  & 17.16  \small{\textbf{(1$^{st}$)}} & $ 1309.41\pm0.02$ & $ -156.94\pm0.02$ &          ...               &            ...              \\
       &    &    &                    ...         &           ...                    &           ...              &           ...                \\  
 51  & --4.27  &   1.36  \small{\textbf{(1$^{st}$)}} & $ 1279.61\pm0.15$ & $ -176.55\pm0.10$ & ... & ...   \\ 
 52  & --4.32  &   3.19  \small{\textbf{(1$^{st}$)}} & $ 1287.24\pm0.05$ & $ -163.20\pm0.04$ & $ -0.65\pm0.26$ & $ -0.03\pm0.17$   \\

\multicolumn{4}{l}{\textbf{Cluster\,V:  mean LSR velocity of --4.54\,km\,s$^{-1}$}} & &  \\
 53  &  --4.36  &  1.55  \small{\textbf{(1$^{st}$)}} & $  111.09\pm0.10$ & $  680.53\pm0.18$ & $  0.38\pm0.18$ & $ -1.31\pm0.35$   \\ 
 54  &  --4.36  &  1.10  \small{\textbf{(1$^{st}$)}} & $  118.36\pm0.19$ & $  665.06\pm0.32$ & ... & ...   \\
 55  &  --4.41  &  1.30  \small{\textbf{(1$^{st}$)}} & $  102.88\pm0.10$ & $  694.45\pm0.16$ & ... & ...  \\
 56  & --4.41  &  0.24  \small{\textbf{(3$^{rd}$)}} & $   94.65\pm0.11$ & $  703.51\pm0.18$ & ... & ...   \\ 
 57  &  --4.54  &  0.26  \small{\textbf{(1$^{st}$)}} & $  131.94\pm0.17$ & $  518.21\pm0.17$ & $  1.07\pm0.55$ & $ -3.35\pm0.56$   \\ 
 58  &  --4.58  &  0.90  \small{\textbf{(1$^{st}$)}} & $  122.06\pm0.08$ & $  527.51\pm0.08$ & $  0.59\pm0.19$ & $ -1.47\pm0.20$   \\ 
 59  &  --4.67  &  0.51  \small{\textbf{(1$^{st}$)}} & $  102.48\pm0.14$ & $  546.56\pm0.12$ & $  0.27\pm0.41$ & $ -2.27\pm0.36$   \\ 
 60  &  --4.71  &  1.78  \small{\textbf{(1$^{st}$)}} & $   92.14\pm0.03$ & $  556.99\pm0.04$ & $  1.32\pm0.11$ & $ -2.03\pm0.11$   \\ 
 61  &  --4.80  &  0.29  \small{\textbf{(1$^{st}$)}} & $   86.64\pm0.10$ & $  551.32\pm0.09$ & $  1.75\pm0.38$ & $ -0.96\pm0.35$  \\ 
 62  &  --4.80  &  0.29  \small{\textbf{(1$^{st}$)}} & $   73.17\pm0.17$ & $  573.26\pm0.16$ & $  1.57\pm0.48$ & $ -2.21\pm0.45$   \\ 
 63  &  --4.93  &  0.18  \small{\textbf{(1$^{st}$)}} & $  -27.66\pm0.16$ & $  551.60\pm0.14$ & $  0.34\pm0.59$ & $ -1.07\pm0.43$  \\ 
 64  &  --4.93  &  0.07  \small{\textbf{(1$^{st}$)}} & $  -35.71\pm0.29$ & $  554.01\pm0.37$ & ... & ...   \\

 & & & & & &  \\ 

\end{longtable}
}

\end{document}